\documentclass{article}

\usepackage{Arxiv}
\usepackage[utf8]{inputenc} 
\usepackage[T1]{fontenc}    
\usepackage{url}            
\usepackage{booktabs}       
\usepackage{amsfonts}       
\usepackage{nicefrac}       
\usepackage{fancyhdr}       
\usepackage{graphicx}       
\graphicspath{{media/}}     
\usepackage[linesnumbered,lined,ruled]{algorithm2e}
\usepackage{array}
\usepackage{amsmath}
\usepackage{amssymb}
\usepackage{subcaption}

\usepackage{multirow}
\usepackage{booktabs}

\title{\textbf{Application of Data Encryption in Chinese Named Entity Recognition}}

\author{
  Kaifang Long, Jikun Dong, Shengyu Fan, Yanfang Geng,\\ \textbf{Yang Cao, Han Zhao, Hui Yu, Weizhi Xu}\\
  School of Information Science and Engineering, Shandong Normal University,\\ Jinan 250358, China \\
  School of Business, Shandong Normal University, Jinan 250358, China \\
  1642445417@qq.com, xuweizhi@sdnu.edu.cn
}

\begin{document}
\maketitle

\begin{abstract}
Recently, with the continuous development of deep learning, the performance of named entity recognition tasks has been dramatically improved. However, the privacy and the confidentiality of data in some specific fields, such as biomedical and military, cause insufficient data to support the training of deep neural networks. In this paper, we propose an encryption learning framework to address the problems of data leakage and inconvenient disclosure of sensitive data in certain domains. We introduce multiple encryption algorithms to encrypt training data in the named entity recognition task for the first time. In other words, we train the deep neural network using the encrypted data. We conduct experiments on six Chinese datasets, three of which are constructed by ourselves. The experimental results show that the encryption method achieves satisfactory results. The performance of some models trained with encrypted data even exceeds the performance of the unencrypted method, which verifies the effectiveness of the introduced encryption method and solves the problem of data leakage to a certain extent.
\end{abstract}


\section{Introduction}
\label{sec:introduction}
Deep learning achieves state-of-the-art performance for many natural language processing (NLP) tasks such as named entity recognition (NER). However, many datasets used for training the deep learning model contain sensitive information. For example, datasets in the biomedical field usually consist of electronic medical records, which generally include identification information, disease information and treatment plans for patients. Many people or groups worry about the leak of their private data and are unwilling to disclose their data. Therefore, data absence is severe in many fields for model training. Most enterprises have problems such as a limited amount of data and poor data quality and cannot support the realization of artificial intelligence (AI) technology. In order to solve the above problem, federated learning \cite{mcmahan2017communication,kairouz2021advances,hitaj2017deep,liu2020secure,li2019fedmd,he2020group} came into being.  

However, there are some shortcomings in federated learning. For example, 1) The data is not centralized. 2) It is impossible to foresee and avoid unstable connections between networks. Once the network is disconnected, the learning process will time out or exit abnormally. 3) Since the federal learning system requires multi-party collaboration, there are problems such as users cannot change models at will, the training speed is slow, and the model training requires high hardware configuration.  4) Although we can see that federated learning has been applied in some practical business scenarios, this technology is still far from entering the stage of large-scale implementation. To tackle the above problems, we propose a deep learning framework based on multiple data encryptions for the NER task. The proposed framework can overcome some shortcomings of federal learning and avoid privacy leakage to some degree.

NER is an important task in NLP, which identifies useful entities in the unstructured text, such as person name, place name, organization name, time, etc. The performance of NER seriously affects its downstream tasks, including relation extraction \cite{miwa2016end}, knowledge graph \cite{chen2021fede}, question answering \cite{diefenbach2018core}, etc. With the emergence of dynamic word embeddings such as BERT \cite{devlin2018bert}, the application of recurrent neural network (RNN) suitable for time series modeling and conditional random field (CRF) with label constraints improves the performance of English NER \cite{wang2020automated,yu2020named}.

Compared with English NER, Chinese NER faces more difficulties. 1) There is no natural space as a character-to-character separator in Chinese sequences. 2) Because of the different lengths of entities in Chinese sentences, it is difficult to determine the boundaries of entities. Even if Chinese word segmentation tools can obtain the boundaries of some entities with different granularities, the error propagation caused by word segmentation cannot be avoided. In the wake of the development of pre-trained language models and character-word-based models like Lattice LSTM \cite{zhang2018chinese}, the performance is satisfactory on resource-rich Chinese datasets \cite{zhang2018chinese,weischedel2011ontonotes,levow2006third,peng2015named,he2017f} such as MSRA and Resume. 

As an important international language, Chinese is unique in many ways. Chinese language, culture, and history are receiving more and more international attention and study, but there is a shortage of datasets about Chinese history in the Chinese NER field. The lack of data and applications poses challenges for scholars studying Chinese history for the first time and the construction of knowledge graphs on Chinese history. Based on the above, we construct a new Chinese historical dataset. 

Meanwhile, we conduct experiments using multiple encryption methods on six datasets in biomedicine, news and history domains. The experiments show that the performance of encrypted data is satisfactory. This proves that our approach ensures the accuracy of the deep learning models and prevents data leakage to a certain extent. Our main contributions are as follows.
\begin{itemize}
\item We introduce the hash algorithms and the ciphertext policy attribute-based encryption (CP-ABE) to NER for the first time. Experiments on six datasets show that our proposed multi-encryption strategy can ensure the performance of the model and protect the data to some degree.
\item We have an interesting finding that the performance does not degrade significantly when encrypting the training data for the NER task.
\item We release a new historical dataset for Chinese named entity recognition. It provides a foundation for recognizing identities from historical documents and building knowledge graphs.
\end{itemize}

\section{Related Work}
\subsection{Named Entity Recognition}
Named entity recognition aims to quickly extract entity information of specific types from complex natural language texts, which provides the foundation for information extraction to generate structured data. The earliest methods of NER are rule-based methods, lexicon-based methods and statistical machine learning-based methods, such as the Support Vector Machine Model (SVM) \cite{habib2007language}, Hidden Markov Model (HMM) \cite{zhou2002named}, Conditional Random Field (CRF) \cite{sobhana2010conditional}, and so on. However, these methods suffer from feature engineering and fail to achieve the desired results for NER. 

With the development of artificial intelligence, numerous neural network methods have been applied and achieved good results for named entity recognition. At that time, researchers mainly used LSTM-CRF or CNN-CRF models to encode and decode the input features \cite{huang2015bidirectional,lample2016neural,ma2016end,chiu2016named,strubell2017fast,wu2019neural,yang2018ncrf++}. But there are some issues with these methods. 1) The character-based methods do not use dictionary information. 2) The word-based method will cause error propagation due to word segmentation errors.

Accordingly, Zhang and Yang proposed a character-word-based hybrid model in 2018. This model can integrate word information into the character of the input sequence, which further improves the performance of the NER task. Therefore, many researchers have derived a large number of models based on the approach . For example, \cite{liu2019encoding} proposed the WC-LSTM model, \cite{gui2019cnn} proposed the LR-CNN model, and \cite{ma2020simplify} proposed the Softword method. \cite{ding2019neural,sui2019leverage,gui2019lexicon} used graph neural networks to improve NER performance. It is well known that NER tasks rely heavily on word embeddings. The quality of word embeddings can determine the performance of the model. The emergence of dynamic word embeddings such as BERT pushed the performance of NER to a new level. For instance, \cite{li2020flat} and \cite{mengge2020porous} proposed a FLAT model and a Porous Lattice Transformer Encoder based on Lattice LSTM, respectively.

\subsection{Data protection}
Named entity recognition introduces data protection \cite{sivakumar2017enhanced,gai2017privacy,mivule2017data} to protect the private information contained in a dataset from being widely read and leaked. However, data leakage is the intentional or unintentional disclosure or loss of data to untrustworthy third parties. This paper focuses on protecting private data using data desensitization and access control strategies to prevent data leakage. Data desensitization is a crucial method to remove the sensitivity of personal privacy data and minimize the risk of data leakage. It is mainly used to process the original data utilizing data replacement, data randomization, data encryption, and hash transformation. We performed a hash transformation on the NER dataset using data desensitization. Second, access control generally means that the data provider sets up security rules or policies. Users can obtain and decrypt the encrypted data according to their permissions or attributes.

\begin{figure*}[h]
\centering
\includegraphics[width=0.85\textwidth]{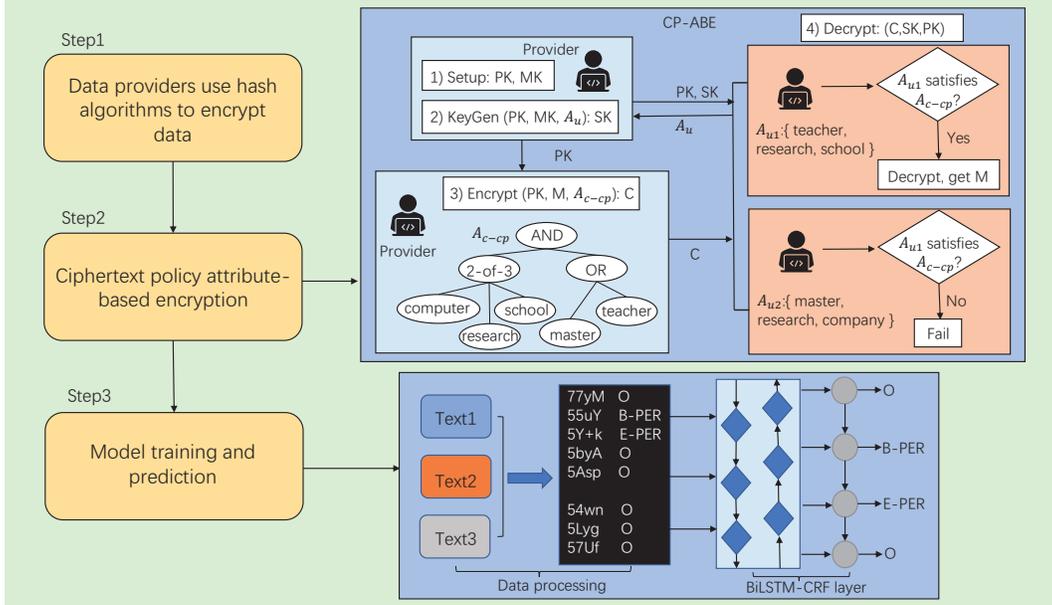}
\caption{ The overall architecture of the proposed method.}
\label{Fig:1}    
\end{figure*}
\section{Method}
To avoid widespread reading and leakage of data, and yet ensure that users can build machine learning models according to their needs. We propose an NER framework based on multi-encrypted data to solve the above problem. As shown in Figure 1, the framework consists of three steps. The first step is that the data provider encrypts the original data using hash algorithms. The second step is that we use Ciphertext policy attribute-based encryption (CP-ABE) based on hash encryption to achieve double data protection to solve the problem of illegal users accessing the data. The third step is that the training and prediction of the model after the legitimate users obtain the data.

\subsection{Data encryption}
Hash functions are used to encrypt data in our model framework in Figure 1. To verify the authenticity and validity of encrypted data training and encrypt the same data using more than one encryption method, we also introduce Serial Cipher and Base64 methods besides hash functions.

Serial Cipher  \cite{stinson2005cryptography}, also called Stream Cipher, is one of the symmetric cryptographic algorithms. It encrypts the plaintext by adding the key. Since serial encryption has the advantages of simple implementation and fast encryption speed, we take it as an optional method for data encryption. Meanwhile, the Caesar Cipher and Affine Cipher can be consistent with the Serial Cipher by adjusting the coefficients. We grouped the three encryption algorithms into one category.

As we all know, Base64 \cite{josefsson2006base16} is the most common encoding method for transmission in the network, which is used to transmit 8-bit byte codes. Although it violates the principle of non-disclosure of encryption keys, Base64 encoding can process text data due to its unreadable advantages. Therefore, the method provides a way for verifying the validity of the encryption method and performing multiple encryptions.

MD5 message-digest algorithm and Has256 algorithm \cite{rivest1992md5} are hash functions widely used in computer security. They provide the ability to transform data with arbitrary length to a fixed value. At the same time, the MD5 algorithm and the Has256 algorithm are irreversible, which can effectively ensure the security of the dataset. In the experiment, to realize the feasibility of multi-encryption of the same data, we also use the Base64 algorithm on top of the Has256 algorithm for encryption.

\subsection{CP-ABE}
Traditional attribute-based encryption (ABE) \cite{goyal2006attribute} systems describe the ciphertext by attributes and embed the policy into the user's key. Where attributes are characteristics of things or information files, and policy is a logical expression composed of attributes and relationships. CP-ABE \cite{bethencourt2007ciphertext,wan2011hasbe} uses attributes to portray the user's eligibility. The data provider makes the ciphertext acquisition policy to decide who can decrypt the ciphertext. In other words, the attributes are embedded in the key, and the policy is embedded in the ciphertext.

As shown in the upper right of Figure 1, firstly, the data provider initializes a public key (PK) and a master key (MK). Secondly, the public key, the master key, and the user's attribute set (Au) generate a private key (SK). Then, the data provider constructs an access control policy (Ac-cp) based on the user's attributes. The public key, the plaintext M ( M indicates the data encrypted with the hash algorithm in the figure), can generate the ciphertext (C) through the access control policy. Finally, the user could decrypt the data based on the public and private keys. If the user attribute is legitimate, the ciphertext can be decoded successfully; otherwise, it cannot.

\subsection{Model training and prediction}
As shown in the bottom right of Figure 1, we describe how the user trains the model and predicts with the model on the encrypted dataset. First, the legitimate user follows the CP-ABE rule and decodes the data to obtain three data files using the ciphertext, public key, and private key provided by the data provider. Text1 represents the ciphertext encrypted with the hash algorithm, Text2 represents the label text corresponding to the ciphertext, and Text3 is the length of each sequence in the ciphertext. According to the three texts, users can get the training data completely. If users need to test the model's performance with their data, users could encrypt the data using the same hashing algorithm in Text1 before prediction.

With the continuous exploration of researchers, neural networks have developed rapidly. Li et al. significantly improved the performance of biomedical named entity recognition by using recurrent neural networks \cite{li2020character}. Meanwhile, applying deep neural networks such as convolutional neural networks, self-attention mechanisms, and transformers \cite{vaswani2017attention,cao2018adversarial} has effectively promoted the development of NER. Therefore, the user can build models based on the above deep neural networks after obtaining encrypted data. Here, we use the classical model BiLSTM-CRF as a benchmark. BiLSTM comprises a forget gate, an input gate, and an output gate. These three gate mechanisms interact with each other and update the cell state. The specific formulas are as follows.
\begin{equation}
F^T=\sigma \, (W^F_\alpha  X^T + W^F_\beta  H^{T-1} +B_F)    \label{equat:8} 
\end{equation}
\begin{equation}
I^T=\sigma \,(W^I_\alpha X^T+ W^I_\beta H^{T-1} +B_I)  \label{equat:9}
\end{equation}
\begin{equation}
O^T=\sigma \,(W^O_\alpha X^T+W^O_\beta H^{T-1} +B_O)   \label{equat:10}
\end{equation}
\begin{equation}
\tilde{C}^T=Tanh \,(W^C_\alpha X^T+W^C_\beta H^{T-1} +B_C)    \label{equat:11}
\end{equation}
\begin{equation}
C^T= F^T\odot \, C^{T-1}+I^T \odot \,\tilde{C}^T \label{equat:12}
\end{equation}
\begin{equation}
H^T =O^T \odot \, Tanh \,(C^T)\label{equat:13}
\end{equation}

Here, $F^T$ represents the information that the cell state will forget. $I^T$ and $O^T$ denote the input and output gates, respectively. $\tilde{C}^T$ denotes the current cell state and $C^T $ denotes the final cell state. Where $W$ is the hyperparameter and $H$ denotes the output of the hidden state.
\begin{figure}[h]
\centering
\includegraphics[width=0.45\textwidth]{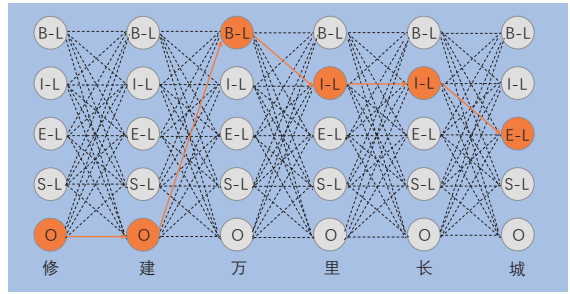}
\caption{ Label state transitions of conditional random fields in named entity recognition.}
\label{Fig:2}    
\end{figure}

We use a conditional random field \cite{forney1973viterbi,lafferty2001conditional,lin2020enhanced} to constrain the transfer between tags after the encoding layer. As shown in Figure 2, B-L denotes the start of a location, I-L denotes the middle of a location, and E-L denotes the end of a location. S-L indicates that a single entity constitutes a location, and O is a non-entity. According to the figure, we perform the following constraints. O is impossible to transfer into I-L and E-L. B-L is impossible to transfer into B-L, S-L, and O.  I-L is impossible to transfer into B-L, S-L, and O.  E-L is impossible to transfer into E-L and I-L.


\section{Datasets}
This paper focuses on comparing model performance before and after data encryption. We use six Chinese NER datasets to verify the authenticity and effectiveness of the experiments, including CCKS2017, Resume, MSRA,  and History. The number of sentences and words in each dataset is shown in Table 1.
\begin{table}[h]
\centering
\setlength{\tabcolsep}{1mm}{
\begin{tabular}{ccccc}		
\hline
Datasets&Type&Train&Dev&Test\\
\hline
\multirow{2}*{Resume} & Char & 124.4K &13.9K & 15.1K \\
~  & Sentence &3.8K& 0.46K & 0.48K\\
\hline
\multirow{2}*{MSRA} & Char & 2169.9K & -- & 172.6K \\
~  &Sentence & 46.4K& -- & 4.4K\\
\hline
\multirow{2}*{CCKS2017} &Char & 200.0K & 31.8K & 33.6K \\
~  & Sentence &5.9K& 0.82K & 1.09K\\
\hline
\multirow{2}*{History} & Char & 289.1K & 30.9K & 0.97K \\
~  & Sentence & 8.9K& 0.97K & 0.81K\\
\hline
\end{tabular}}
\caption{\label{Tab01}
Statistics of datasets}
\end{table}

\begin{table}[h]
\centering
\label{Tab02}
\setlength{\tabcolsep}{1.6mm}{
\begin{tabular}{|cccccc|}
\hline
\multicolumn{1}{|c|}{--}    & \multicolumn{1}{c|}{ORGARM}  & \multicolumn{1}{c|}{LOC}  & \multicolumn{1}{c|}{DAT}  & \multicolumn{1}{c|}{ORG}    & PER \\ \hline

\multicolumn{1}{|c|}{Train} & \multicolumn{1}{c|}{1202} & \multicolumn{1}{c|}{4231} & \multicolumn{1}{c|}{1510} & \multicolumn{1}{c|}{3934}   & 8618   \\ \hline

\multicolumn{1}{|c|}{Dev}   & \multicolumn{1}{c|}{215}  & \multicolumn{1}{c|}{578}  & \multicolumn{1}{c|}{179}  & \multicolumn{1}{c|}{426}    & 829    \\ \hline

\multicolumn{1}{|c|}{Test}  & \multicolumn{1}{c|}{212}  & \multicolumn{1}{c|}{603}  & \multicolumn{1}{c|}{311}  & \multicolumn{1}{c|}{396}    & 507    \\ \hline 

\multicolumn{1}{|c|}{--}    & \multicolumn{1}{c|}{LOCPER}  & \multicolumn{1}{c|}{EVE}  & \multicolumn{1}{c|}{POS}  & \multicolumn{1}{c|}{APP} & --     \\ \hline

\multicolumn{1}{|c|}{Train} & \multicolumn{1}{c|}{231}  & \multicolumn{1}{c|}{383}  & \multicolumn{1}{c|}{3169} & \multicolumn{1}{c|}{834}    & --     \\ \hline

\multicolumn{1}{|c|}{Dev}   & \multicolumn{1}{c|}{34}  & \multicolumn{1}{c|}{19}   & \multicolumn{1}{c|}{441}  & \multicolumn{1}{c|}{111}     & --     \\ \hline

\multicolumn{1}{|c|}{Test}  & \multicolumn{1}{c|}{32}  & \multicolumn{1}{c|}{64}   & \multicolumn{1}{c|}{216}  & \multicolumn{1}{c|}{118}     & --     \\ \hline
\end{tabular}}
\caption{\label{Tab02}
The number of nine entities on the train, dev, and test sets.}
\end{table}
CCKS2017 is a clinical medicine NER dataset released by China Conference on Knowledge Graph and Semantic Computing. Since CCKS2017 only provides a relatively large-scale dataset, we divided the dataset to test set, development set, and training set. Resume and MSRA are from social media and news, and the History dataset comes from the field of Chinese history.

\begin{table*}[ht]
\scriptsize
\centering
\setlength{\tabcolsep}{3mm}{
\begin{tabular}{cccccccccc}
\toprule
\multirow{2}{*}{Model} & \multicolumn{3}{c}{CCKS2017} & \multicolumn{3}{c}{Resume} & \multicolumn{3}{c}{MSRA} \\
\cmidrule(r){2-4} \cmidrule(r){5-7} \cmidrule(r){8-10}
&  $P$      &  $R$   &   $F1$
&  $P$      &  $R$   &   $F1$
&  $P$      &  $R$   &   $F1$ \\
\midrule
LSTM-CRF             &88.45                         &87.35                    & 87.90                   & \textbf{93.73} & 93.44 &93.58          &89.52           & \textbf{87.41}         &88.45         \\

+ Serial Cipher             &\textbf{89.25}                          &87.13                    &\textbf{88.18}                  & 93.53           & 93.13         &93.33          &89.10           & 87.25          & 88.16 \\

+Base64             &88.66                         & 87.11                    & 87.88                  &93.68 & 93.62           & \textbf{93.65}          &\textbf{89.74}          & 87.34          &\textbf{88.53}         \\

+MD5              &89.21                          & 86.96                  &88.07              &93.46          & \textbf{93.80} &93.63         &89.68           &87.18        & 88.41            \\
+Has256-Base64              &87.74                          & \textbf{87.48}                    &87.61                 & 93.50 & 93.56 & 93.53         &88.14           & 86.19        &87.15          \\
\midrule

WC-LSTM             &88.96                          & 87.33                    & 88.14                   & 95.14           & \textbf{94.79}          &94.96        & \textbf{93.67}         &\textbf{92.20}           &\textbf{92.93}       \\

+ Serial Cipher              &\textbf{90.55}                          & 86.14                    & 88.29                   & \textbf{95.36}           & 94.66          & \textbf{95.01}          & 93.66          & 92.02         & 92.83        \\

+Base64              &89.59                          &\textbf{87.38}                   & \textbf{88.47}                   & 93.64 & 93.87           &93.75         &89.33          &87.58         &88.45           \\

+MD5            &89.43                         & 87.33                 &88.37                  &93.65 &94.05 & 93.85 &88.81           &86.92        & 87.86          \\

+Has256-Base64              &88.96                          & 87.21                    &88.08                  & 93.57 & 93.74 &93.66         &90.23          &87.34       & 88.76           \\
\midrule

Multi-digraph             &89.50                          &88.40                  &88.94                 & 94.62           &94.97          &94.79         &90.82          &\textbf{91.20}       &91.01        \\

+ Serial Cipher              &88.43                       &86.01                    &87.20                   & \textbf{95.04}           & \textbf{95.28}        &\textbf{95.16}          &88.21          &85.73        & 86.95        \\

+Base64              &\textbf{89.66}                         &88.86                   &\textbf{89.26}                   & 94.44 &94.91 &94.68        & 91.30           &90.68        &90.99          \\

+MD5             &89.22                       & 88.94                   &89.08                 &94.37         & 94.66         &94.52      &90.54           &87.19        &88.83           \\

+Has256-Base64              &88.73                         & \textbf{89.18}                    &88.95                & 94.74 & 95.03 &94.89         &\textbf{91.38}           &90.86         &\textbf{91.12}          \\
\midrule

SoftLexicon            &89.67                         & 87.23                 &88.43 &95.30 &\textbf{95.77} & \textbf{95.53} &\textbf{93.72}         &\textbf{91.88}          &\textbf{92.79}        \\

+ Serial Cipher             &90.08                          &\textbf{88.34}                    &\textbf{89.20}                  & 95.48           &94.66        &95.07          & 88.42           &84.32           &86.32         \\

+Base64             &89.74                          & 86.89                  & 88.29                   &94.25 & 93.50 & 93.87         & 88.44           &84.63           &86.49           \\

+MD5            &\textbf{90.52}                         & 86.98                    &88.72                  & \textbf{95.50}           &94.97 & 95.23          &88.23      &84.52          & 86.33        \\

+Has256-Base64              &90.46                         & 86.42                   &88.39                & 95.26 & 94.97 &95.12         &87.83           &84.68         &86.23           \\
\toprule
\end{tabular}}
\caption{\label{Tab02}
Performance on three public datasets}
\end{table*}


\begin{table*}[htb]
\scriptsize
\centering
\setlength{\tabcolsep}{3mm}{
\begin{tabular}{cccccccccc}
\toprule
\multirow{2}{*}{Model} & \multicolumn{3}{c}{History-9types} & \multicolumn{3}{c}{History-3types} & \multicolumn{3}{c}{History-2types} \\
\cmidrule(r){2-4} \cmidrule(r){5-7} \cmidrule(r){8-10}
&  $P$      &  $R$   &   $F1$
&  $P$      &  $R$   &   $F1$
&  $P$      &  $R$   &   $F1$ \\
\midrule
LSTM-CRF             &76.01                         &\textbf{60.68}                    & \textbf{67.48}                   & 76.15           & 50.45 & \textbf{60.69}          & 72.40           & \textbf{60.58}           & \textbf{65.96}          \\

+ Serial Cipher             &76.04                          &60.27                    &67.24                   & \textbf{76.94}           & 49.36          & 60.09          &71.32           & 52.97           & 60.79          \\

+Base64             &76.06                          & 60.59                    & 67.45                   &73.60           & 51.85           & 60.84          &72.14           & 60.17           &65.61          \\

+MD5              &\textbf{76.88}                          & 59.90                    &67.34                  & 72.18          & \textbf{52.23} & 60.61         & \textbf{73.18}           & 58.51         & 65.03            \\
+Has256-Base64              &77.20                          & 59.90                    &67.46                  & 74.25 & 50.32 & 59.98         & 75.00           & 57.68         & 65.21            \\
\midrule

WC-LSTM             &\textbf{82.43}                          & \textbf{68.48}                    & \textbf{74.81}                   & \textbf{83.19}           & 61.15           & 70.48        & 82.45           &62.38           &71.02          \\

+ Serial Cipher              &82.30                          & 68.08                    & 74.52                   & 81.99           & \textbf{62.04}          & \textbf{70.63}          & \textbf{84.73}          & \textbf{62.93}         & \textbf{72.22}          \\

+Base64              &77.84                          &61.12                    & 68.47                   & 73.02           & 51.72           &60.55          &73.47           &61.27           & 66.82           \\

+MD5            &76.27                          & 62.10                   & 68.46                   & 72.99           &51.94 & 60.71 &74.15           &60.30          & 66.51           \\
+Has256-Base64              &76.67                          & 61.61                    &68.32                  & 75.57 & 50.45 &60.50         & 72.50           &61.27         & 66.42            \\
\midrule

Multi-digraph             &75.31                          &71.70                  &73.46                  & \textbf{76.14}           & 59.36           &\textbf{66.71}          &75.31           &76.35         &\textbf{75.82}         \\

+ Serial Cipher              &77.24                          & 68.73                    &72.74                   & 69.54           & 61.66         &65.36          &71.54           &\textbf{77.18}         & 74.25         \\

+Base64              &\textbf{78.42}                         &71.37                   &\textbf{74.73}                   & 69.06           &\textbf{62.55} & 65.64        & 74.90           &75.52         & 75.21          \\

+MD5             &75.79                        & \textbf{71.82}                   & 73.75                 & 69.79          & 60.64         & 64.89         & \textbf{76.23}           &70.54         & 73.28            \\
+Has256-Base64              &78.66                          & 63.42                    &70.22                 & 73.87 & 60.51 &66.53         &75.46           &74.00         &74.72            \\
\midrule

SoftLexicon            &\textbf{82.65}                         & \textbf{70.11}                  &\textbf{75.86} &\textbf{82.47} & \textbf{67.13} & \textbf{74.02}       &\textbf{85.71}          &\textbf{73.03}           &\textbf{78.86}        \\

+ Serial Cipher             &80.31                          &66.33                    &72.65                  & 82.23           & 63.06         &71.38          & 82.30           & 71.37           &76.44          \\

+Base64             &73.00                          & 52.99                   & 61.40                   &73.36           & 51.85 & 60.76         & 71.99           &56.15           &63.09           \\

+MD5            &75.26                         & 58.28                    & 65.69                  & 73.01           &54.87 & 62.59          & 79.62      &63.21          & 70.47            \\
+Has256-Base64              &75.24                          & 57.83                    &65.39                 & 74.43 & 53.76 &62.43         &76.97           &64.73         &70.32           \\
\toprule
\end{tabular}}
\caption{\label{Tab02}
Performance on History datasets}
\end{table*}
At present, Chinese has become an important international language. We constructed a brand-new dataset to solve the lack of historical datasets (History in Table 1) in the Chinese NER domain. As shown in Table 2, our dataset consisted of 9 types of labels, including organization name (ORG), place name (LOC), time (DAT), person name (PER), salutation (POS), official position (APP), book name (EVE), army name (ORGARM), and place of belonging (LOCPER). Furthermore, we divided the historical dataset into three categories to identify certain specific entities. The first category has nine types of tags called History-9types. The second category, History-3types, has LOC, APP, and EVE entities. The third category has only PER and POS called History-2types.

\section{Experiments}
\subsection{Baseline Methods}
In this section, we use four models proposed in recent years to verify the effectiveness of encryption algorithms.

\textbf{BiLSTM-CRF.} BiLSTM-CRF was proposed by Lample et al. in 2016. The method is a classical model in named entity recognition. Compared with the traditional machine learning models, it shows a dramatic enhancement in performance.

\textbf{WC-LSTM.} WC-LSTM (2019) is a word-character-based model proposed by Liu et al. for addressing the shortcomings of Lattice LSTM. The model provides four methods that effectively integrate lexicon knowledge into characters.

\textbf{Multi-digraph Model.} Multi-digraph (2019) is a model proposed by modifying the gated graph neural network (GGNN), which can effectively integrate word information into characters.

\textbf{SoftLexicon.} SoftLexicon (2020) is a novel approach to utilizing dictionary information proposed by Ma et al. Its encoding framework is very flexible and can enormously improve the performance of entity recognition.

\subsection{Overall Performances}
Table 3 shows the results on three public datasets, including CCKS2017, Resume, and MSRA. The results of four different models are given in Table3. For each model, we do five experiments. In the first experiment, we train the model and predict with the model using plaintext. In the other four experiments, we train the model and predict with the model using ciphertext, encrypted by Serical Cipher, Base64, MD5 and Has256-Base64 respectively. Has256-Base64 means that the data encrypted first by Has256 and then by Base64. For CCKS2017 and Resume, the experimental results of the four encryption methods show barely any change in performance on the four baseline models when comparing the plaintext training method with the ciphertext training methods. some results obtained with the ciphertext training methods are even better than the plaintext training method. For the MSRA dataset, when training Has256-Base64 encrypted data using Multi-digraph, Precision and F1 value are 0.56\% and 0.11\% higher than the plaintext training method. When training LSTM-CRF, there is almost no degradation in the performance. When training  WC-LSTM and SoftLexicon, the experimental performance degradation may be caused by the under-utilization of lexical knowledge by the encryption methods and the effect of the absence of development sets.

Table 4 shows the results of the historical dataset. As shown in the table, compared to unencrypted, the performance of the four encryption algorithms barely degrades when trained with the LSTM-CRF model. Moreover, some encryption algorithms have better performance than unencrypted ones. For example, Precision is 0.87\% and 0.78\% higher than unencrypted by using the MD5 encryption algorithm on the History-9types and History-2types datasets. Precision is improved by 0.79\% using serial encryption algorithms, and Recall is improved by 1.78\% using MD5 encryption algorithms on the History-3types dataset. When the WC-LSTM model is trained using Serial Cipher encrypted data, Recall is 0.89\% higher than the unencrypted dataset on the History-3types dataset. Precision, Recall, and F1 value are 2.28\%, 0.55\%, and 1.2\% higher than the unencrypted on the History-2types dataset. When trained with the Multi-digraph model, Precision and F1 value are 3.11\% and 1.27\% higher than unencrypted when using the Base64 algorithm on History-9types. When trained with the SoftLexicon model, we find the effectiveness of the encryption algorithm drops significantly. Because the SoftLexicon model breaks the convention of using vocabulary knowledge, it causes poor performance.

\subsection{Analysis}
\subsubsection{Performance analysis}
  Except for the LSTM-CRF model, the other models use vocabulary knowledge. However, the performance of models using vocabulary knowledge degraded on certain datasets. The possible reasons are as follows. 1) The vocabulary knowledge is underutilized after encrypting the data. 2) The embedding of characters and words is possibly affected by the encryption process. 3) Different model architectures also influence the performance. To solve the above problems, we can obtain the word embedding by retraining the model using large-scale encrypted data in the future. At the same time, because of the advantage of dynamic word embedding like BERT, we can also consider training a BERT using encrypted data to improve the performance of NER in the future.  
 
\subsubsection{Security analysis}
In this paper, we use hash functions and CP-ABE to ensure data security. The hash functions have weak collision-resistant, strong collision-resistant, and modification resistance characteristics. The most important feature is that it is irreversible, so users are difficult to decrypt the plaintext, which can ensure data security to a great extent. At the same time, we can further encrypt the data on top of the hash encryption to strengthen the security such as Has256-Base64. Finally, we use CP-ABE to ensure the legitimacy of the user who obtains the ciphertext generated by other encrytion methods. Since CP-ABE is built on the difficulty of computing discrete logarithms, it is challenging to get the data for unauthorized users, proxy servers, etc. Based on the above analysis, we can ensure the feasibility and security of our framework.

\subsection{Relations between entities}
We train a NER model using the Chinese historical dataset to improve the performance of its downstream tasks, which can accurately identify entities in unstructured text. Figure 3 shows a simple knowledge graph constructed after we extracted partial entity types from the text by the model.

The red nodes indicate the Chinese dynasties, such as the Qing Dynasty, Ming Dynasty, and Tang Dynasty. The red edges indicate the substitution between dynasties, such as the Qing Dynasty overthrew the Ming Dynasty and the Tang Dynasty overthrew the Sui Dynasty. The yellow nodes express the capital of each dynasty. As a result of capital city migration, a red node may be connected to multiple yellow nodes. A black node describes the country's leader, and a black edge indicates a father-son or brother relationship. Green nodes are emperors' aliases.

In Figure 3, we can know most of the relationships between the various dynasties, which fully justifies the necessity of constructing a Chinese historical dataset. It paves the way for future research.
\begin{figure}[h]
\centering
\includegraphics[width=0.5\textwidth]{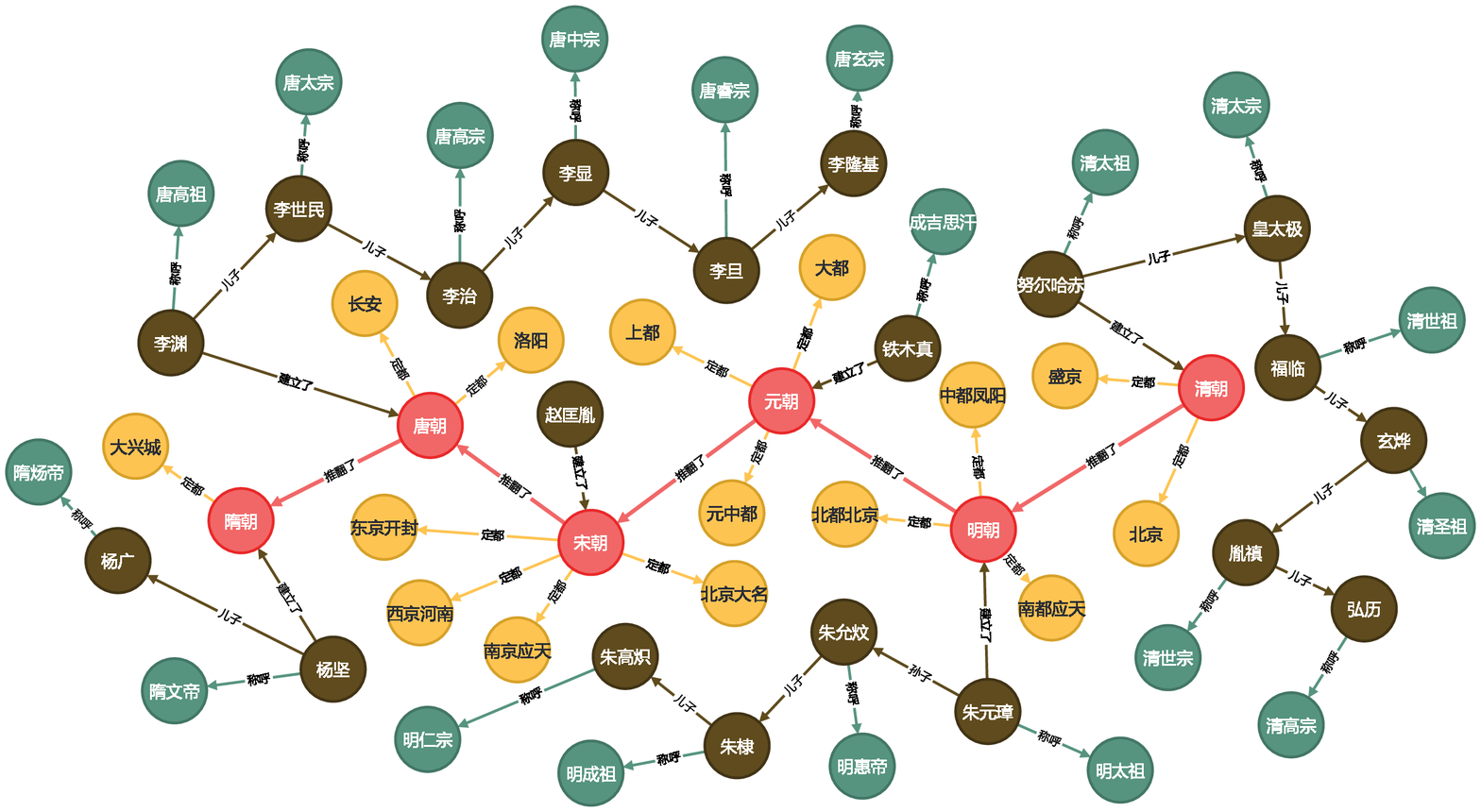}
\caption{ Diagram of the relationship of partial entities in the historical dataset.}
\label{Fig:2}    
\end{figure}

\section{Conclusion}
In this paper, we introduce the hash algorithm and CP-ABE to the named entity recognition task for the first time. This method can effectively solve the problem of data leakage in some fields. We propose a new dataset to solve the problem of the lack of datasets in historical domains. Our experiments on six datasets demonstrate that our method can obtain satisfactory results. Simultaneously, in the future, we can further improve the experimental performance by training BERT using encrypted data.


\bibliographystyle{unsrt}  
\bibliography{references}

\end{document}